\documentclass[letterpaper,fleqn,usenatbib]{mnras}

\usepackage{enumerate}

\def\eg{{\it e.g.}}
\def\etal{{\it et al.}}

\def\NB{{\it NB.}}

\def\Gaia{{\it Gaia\/}}
\def\pmb#1{\setbox0=\hbox{$#1$}%
  \kern-0.25em\copy0\kern-\wd0
  \kern.05em\copy0\kern-\wd0
  \kern-0.025em\raise.0433em\box0}
\def\spmb#1{\setbox1=\hbox{${\scriptstyle #1}$}%
  \kern-0.25em\copy1\kern-\wd1
  \kern.05em\copy1\kern-\wd1
  \kern-0.025em\raise.0433em\box1}

\long\def\Ignore#1{\relax}
\usepackage{color}
\definecolor{red}{rgb}{0.7,0.1,0.1}
\definecolor{blue}{rgb}{0.2,0.2,0.8}
\definecolor{green}{rgb}{0.1,0.6,0.1}

\usepackage{graphicx}	% Including figure files
\title[Spiral saturation and decay]{Spiral instabilities: Mode saturation and decay}

\author[Sellwood \& Carlberg]
          {J. A. Sellwood$^{1}$\thanks{E-mail:sellwood@as.arizona.edu}
and
{R. G. Carlberg$^2$\thanks{E-mail: raymond.carlberg@utoronto.ca}}
\\
$^1$Steward Observatory, University of Arizona, 933 N Cherry Ave,
Tucson AZ 85722, USA \\ 
$^2$Department of Astronomy and Astrophysics, University of Toronto,
ON M5S 3H4, Canada}

\pubyear{2021}

\begin{document}
\label{firstpage}
\pagerange{\pageref{firstpage}--\pageref{lastpage}}
\maketitle

\begin{abstract}
This paper continues a series reporting different aspects of the
behaviour of disc galaxy simulations that support spiral
instabilities.  The focus in this paper is to demonstrate how linear
spiral instabilities saturate and decay, and how the properties of the
disc affect the limiting amplitude of the spirals.  Once again, we
employ idealized models that each possess a single instability that we
follow until it has run its course.  Remarkably, we find a tight
correlation between the growth rate of the mode and its limiting
amplitude, albeit from only six simulations.  We show that non-linear
orbit deflections near corotation cause the mode to saturate, and that
the more time available in a slowly-growing mode creates the critical
deflections at lower amplitude.  We also find that scattering at the
inner Lindblad resonance is insignificant until after the mode has
saturated.  Our objective in this series of papers, which we believe
we have now achieved, has been to develop a convincing and
well-documented account of the physical behaviour of the spiral
patterns that have been observed in simulations by others, and by
ourselves, for many decades.  Understanding the simulations is an
important step towards the greater objective, which is to find
observational evidence from galaxies that could confront the
identified mechanism.
% 208 words
\end{abstract}


%%%%%%%%%%%%%%%%% BODY OF PAPER %%%%%%%%%%%%%%%%%%

\section{Introduction}
\label{sec.intro}
Spiral instabilities have been reported in $N$-body simulations of
rotating disc galaxy models for over 50 years \citep{MPQ, Hohl, HB74}.
This behaviour has revealed that spirals are fundamentally a
collective phenomenon of many-body gravitational dynamics, but a
satisfactory explanation for them has taken many years to emerge.  We
have developed, in a series of papers, an understanding of various
dynamical processes that together furnish a compelling account for the
behaviour in collisionless simulations, which would have been far more
difficult to unravel had they included a gas component.  These papers
present a detailed analysis of simplified models that each build parts
of the case.  Not only does this piecemeal approach make it more
difficult to see the whole picture, but the papers contain much
technical detail that requires careful study.  We therefore give below
an overview in simple terms of the ideas we have developed.  It is to
be hoped that the same mechanism also accounts for the origin of
self-excited spiral patterns in galaxies, but supporting observational
evidence for this, or any other suggested excitation mechanism for
spirals in galaxies, remains flimsy.  See \citet{SM22} for a full
review of both the observed properties of spirals in galaxies and
theoretical ideas to account for them.

Our fundamentally straightforward idea for the origin of spirals in
simulations is as follows.  A disc of particles having a moderate
level of random motion is destabilized by non-smooth features in the
distribution of angular momentum, which are inevitably present in all
particle discs.  A ``groove,'' or deficiency of particles over a
narrow range of angular momentum, gives rise to one, or more, linearly
unstable spiral modes that grow vigorously until saturating at $\la
30\%$ relative overdensity before beginning to decay.  The later
stages of growth, saturation, and early decay enable a steadily
rotating spiral pattern to be detectable for no more than $\sim10$
rotations.  As each spiral mode decays, it interacts with particles
that are in resonance with the spiral, changing their angular momenta.
This localized resonant scattering creates new grooves in the particle
distribution that were not present before.  The new grooves in turn
give rise to fresh instabilities having differing pattern speeds, and
perhaps rotational symmetries, that become new spiral patterns.  The
initial groove to seed this recurrent cycle in a real galaxy could be
caused by resonance scattering as, say, an orbiting mass clump settles
into the disk or by the near passage of a small companion or, in the
unlikely circumstance that neither of these events happen, spiral
disturbances could bootstrap out of the noise \citep{Sell12}.

In general, a disc possesses several spiral modes at any one time,
each having a different pattern speed, rotational symmetry, and
peaking in amplitude at a different time.  The superposition of
multiple modes causes the net spiral appearance to evolve
continuously, and the overall pattern changes unrecognizably within
one orbit period.

Since the scattering of particles by spiral waves is irreversible, the
level of random motion in a collisionless disc rises, making it less
responsive over time and spiral activity fades.  However, it was noted
by \citet{Oort}, and others, that spirals in real galaxies are most
prominent in discs that are forming stars from a moderate fraction of
gas.  \citet{SC84} demonstrated that mimicking the dissipative effects
of gas by adding fresh star particles on near circular orbits could
allow spiral activity to persist ``indefinitely'' (see also
\citealt{CF85, Toom90, ABS} and simulations of galaxy formation,
reviewed \eg\ by \citealt{Voge20}).

The picture sketched in the previous few paragraphs is buttressed by
several detailed papers.  \citet{SK91} developed the theory of
instabilities provoked by narrow features in the angular momentum
density, and presented highly simplified simulations to illustrate
them.  The existence of multiple coherent waves having radially
constant pattern speeds, first hinted at by \citet{Sell89}, was
confirmed by \citet{SC14} in both 2D and 3D simulations having a
thousand times more particles.  That paper also demonstrated that
later instabilities were caused by changes to the distribution
function, since the same disturbances grew afresh when existing
non-axisymmetric features were erased.  \citet{SC19} provided clear
illustrations that scattering at the Lindblad resonances of an
instability created grooves to seed new instabilities.  \citet{SC21}
presented an animation\footnote{Reproduced as \\ {\tt
  http://www.physics.rutgers.edu/$\sim$sellwood/supp\underline{~}material.html},
  and in the supplementary material of \citet{SM22}.} of how two
superposed steady waves can give the impression of a swing-amplified
transient.  The present paper furnishes a study of the mechanism that
arrests the exponential growth of a spiral instability,
illustrates its subsequent decay, and how this behaviour depends on the
mode growth rate.

Other ideas have been proposed for the origin of spirals in galaxies,
which we generally find unconvincing.  \citet{Toom90} and \citet{DVH}
argue that spirals are responses to mass clumps orbiting within the
disc; however, their models support small-scale, multi-arm patterns
that do not resemble the large-scale, predominantly two-arm spirals of
real galaxies \citep[\eg][]{Davis12, Hart2016, Yu18}.  \citet{DB14}
reviewed the idea that spirals result simply from swing-amplification
\citep{Toom81}; not only do proponents of this idea offer no candidate
for what is being amplified, but the empirical evidence they present
results from a misinterpretation of their simulations, as explained in
\S6 of \citet{SC21}.  \citet{BLLT}, in a theory that is not based on
simulations, argue that ``grand design'' spirals result from
slowly-growing bi-symmetric modes that persist for many tens of
rotations in sub-maximal discs.  However the quiescent ``basic state''
they invoke for the disc is at variance with the unrelaxed phase space
in the local disc of the Milky Way that was recently reported by
\citet{STCCR} from an action-angle analysis of \Gaia\ DR2 data
\citep{Gaia18}.  Furthermore, this theory disregards the more vigorous
instabilities having $m>2$ that are expected in sub-maximal discs
\citep{Sell11}.  \citet{Bert14} argues that $N$-body simulations
employ ``too few particles''\footnote{\NB, \citet{Sell12} employed up
  to $N= 5\times 10^8$} to ``properly capture'' the damping of spirals
at Lindblad resonances, despite the repeated demonstration
\citep[\eg][]{SL89, Sell12, SC14, SC19} of scattering of particles at
Lindblad resonances, which is how spiral waves are absorbed and is the
cause of the very instabilities the proponents of the density-wave
theory choose to disregard.

As stated above, the purpose of this paper is to demonstrate the
mechanism that causes a spiral instability to saturate, and to study
its decay.  In order to show this clearly, we employ simulations of a
simplified model for the disc that initially possesses just a single
spiral instability and follow its evolution until its decay has seeded
new instabilities.  In separate experiments, we also study the
saturation of other modes that grow at differing rates.

\section{Technique}
\label{sec.methods}
We employ the same model as \citet{SB02}, which is an otherwise stable
disc model to which an initial groove is inserted to provoke a spiral
instability.  The reason that a groove is destabilizing, which was
first given by \citet{SK91}, is as follows.

\subsection{Mechanism for linear growth}
\label{sec.mechnsm}
If all the particles were to pursue circular orbits then a groove
would be a narrow feature in the surface density.  Infinitesimal,
sinusoidal distortions to the groove edges would bring high density
material into the groove at some azimuths and conversely widen the
groove at other phases of the distortion, creating bulges that
corotate with the material on both edges.  If the bulge on the outer
edge of the groove were to lie ahead of that on the inner edge, then
the gravitational attraction between the two bulges would remove
angular momentum from the outer bulge, causing particles to sink
further towards the disc centre, while those in the inner bulge would
gain and therefore rise outward.  Thus this phase relation between the
two bulges creates a mild instability, even without the supporting
response of the surrounding disc.  The instability is still present in
a disc with random motion, as long as the angular momentum
distribution of the particles has a deficiency over a narrow range.
In this case, however, random motions blur the density variations and
thereby reduce the growth rate.

The growing distortions on the groove edges cause sinusoidal changes
to the surface density around the groove, which therefore provoke a
supporting response from the surrounding disc \citep{JT66}.  The
vigour of the large-scale supporting response, which is directly
related to swing-amplification \citep{GLB65, JT66, Toom81, SM22},
varies with the disc properties, and generally strongly enhances the
mild growth rate of the groove instability that would be expected in
its absence \citep{SK91}.

\subsection{Adopted model}
\label{sec.grooved}
Our disc model is a half-mass Mestel disc having an initial $Q=1.5$,
with the distribution function (DF) given by \citet{BT08}.  In order
to limit its radial extent, the disc is tapered by a central cutout
(index $\nu=4$) and outer taper (index $\mu=6$, centered on $R=15$),
that are both gentle enough so as not be destabilizing.  The surface
density of the remaining disc is close to its untapered value over the
range $2 \la R \la 10$.  This model was predicted by \citet{Toom81},
and confirmed in simulations by \citet{Sell12}, to have no linear
instabilities whatsoever.

In order to provoke a spiral instability, we add a groove to this
doubly tapered disc, $f_0$.  The groove has a Lorentzian function
form, so our final DF is
\begin{equation}
f(E,L) = f_0(E,L) \left[ 1 + {\beta w_L^2 \over (L-L_*)^2 + w_L^2} \right].
\label{eq.DF}
\end{equation}
The first simulation we present in this paper uses values for the
groove parameters adopted in \citet{SB02} namely: depth $\beta=-0.4$,
width $w_L=0.2$, and central $L_*=6.5$.  In \S\ref{sec.variants} we
report the effects of changing the $Q$ of the Mestel disc and the
parameters of the groove.

The tapers and groove remove mass from the disc, but we ensure that
the central attraction is $a_R=-V_0^2/R$ at all radii, implying a
rigid halo and preserving the disc equilibrium.  \citet{Sell21}
verified that spiral dynamics is very little affected when a live halo
is substituted for a rigid one.

Here, and throughout the paper, we use units such that $G=V_0=R_0=1$,
where $V_0$ is the circular speed of the Mestel disc and $R_0$ is the
central radius of the inner cut out.  Our unit of time is therefore
$\tau_{\rm dyn}=R_0/V_0$.

We select particles from this DF (eq.~\ref{eq.DF}) using the method
described in the appendix of \citet{DS00}, and place copies of each
particle regularly around a circle to create a quiet start
\citep{Sell83}.

\begin{table}
\caption{Default numerical parameters} % run5203
\label{tab.DBHpars}
\begin{tabular}{@{}ll}
Grid points in $(r, \phi, z)$ & 230 $\times$ 256 \\% $\times$ 125 \\
Grid scaling & $R_0= 10$ grid units \\
%Vertical spacing & $\delta z = 0.04R_0$ \\
Active sectoral harmonics & see text \\
Softening length & $R_0/20$ \\
Number of massive particles & $6 \times 10^6$ \\
Number of test particles & $6.8 \times 10^3$ \\
Time-step & $0.08R_0/V_0$ \\
Time step zones & 5 \\
\end{tabular}
\end{table}

\subsection{{\sc Galaxy} code}
The particles in our simulations move over a 2D polar mesh.  The
gravitational attractions between the particles are calculated at grid
points and interpolated to the position of each particle.  A full
description of our numerical procedures is given in the on-line manual
\citep{Sell14} and the source code is available for download.  Table
\ref{tab.DBHpars} gives the values of the numerical parameters adopted
for most simulations presented in this paper, which indicates that we
typicaly employ ten times more particles than did \citet{SB02}.  We
note where these values were changed.

As usual, we measure non-axisymmetric distortions of the distribution
of the $N$ massive particles using an expansion in logarithmic spirals:
\begin{equation}
A(m,\gamma,t) = {1 \over N}\sum_{j=1}^N \exp[im(\phi_j + \tan\gamma \ln R_j)],
\label{eq.logspi}
\end{equation}
where $(R_j,\phi_j)$ are the polar coordinates of the $j$th particle
at time $t$, $m$ is the sectoral harmonic, and $\gamma$ is the
(radially constant) angle of the spiral component to the radius
vector, which is the complement to the pitch angle.

For this study, we also introduce a set of 34 rings, each having 200
equally spaced test particles, spanning the radial range of the
principal mode $1.5 \leq R \leq 12.5$.  These massless particles have
initially circular orbits, but move in response to the central
attraction and the disturbance forces from the massive disc particles.
We find that their behaviour provides valuable additional information
about the response of the disc to the growing mode.

\section{Results}
\subsection{Imposed bisymmetry}
\label{sec.res1}
\begin{figure}
\includegraphics[width=\hsize,angle=0]{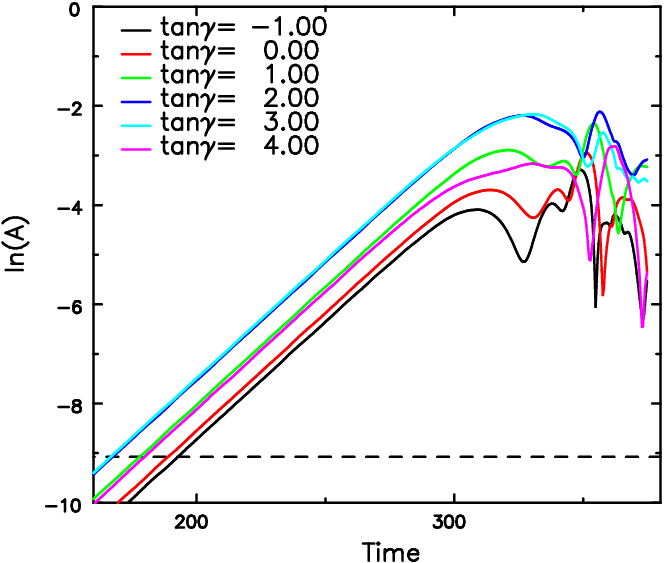}
% jasmine:/data/sellwood/5200/5205/amplot.s
\caption{The second half of the amplitude evolution of several
  components of the logarithmic spiral transforms
  (eq.~\ref{eq.logspi}). The horizontal dashed line marks the
  amplitude expected if the particles were randomly distributed.  The
  quiet start ensures that the initial amplitude is much lower at
  first, and each coefficient grows exponentially until the
  instability saturates at $t \sim 300$.}
\label{fig.amplot}
\end{figure}

As reported by \citet{SB02}, the model described in
\S\ref{sec.grooved} with disturbance forces confined to $m=2$,
possessed a single strong instability.  Note we employed $N=60$
million particles for this first simulation.  Figure~\ref{fig.amplot}
presents the amplitude evolution of six of the $m=2$ logarithmic
spiral coefficients (eq.~\ref{eq.logspi}) over the second half of the
evolution; the first half is dominated by exponential growth from low
amplitude.  The lines for different values of $\tan\gamma$ are
parallel to $t \simeq 280$, indicating the disturbance has a fixed
shape while growing exponentially, as expected for a normal mode.  The
trailing spiral terms $\tan\gamma = 2$ and 3 have the largest
amplitude while the leading $\tan\gamma = -1$ is the smallest shown,
and this leading/trailing bias arises because the disc response to the
growing bisymmetric distortion around the groove \citep{SK91} is
closely related to swing-amplification, as explained in
\S\ref{sec.mechnsm} above.  Clearly, the instability saturates around
$t \sim 300$, after which the shape of the disturbance changes.

The quiet start enabled us to fit a mode \citep{SA86} to the
simulation data over the whole growing phase.  The best fit
eigenfrequency of the mode is $\omega = m\Omega_p + i\beta =
0.282\pm0.000 + (0.046\pm0.001)i$, with the uncertainties indicating
the full range of values from different time intervals and choices of
fitted coefficients.  Fits over time ranges to $t=280$ are excellent
and leave tiny residuals after subtracting this single mode from the
data.  This measured frequency is the same as that reported by
\citet{SB02}, but the larger number of particles employed here gives
the mode a lower seed amplitude and the time of mode saturation is
therefore about 30 dynamical times later.

The radii of the principal resonances for circular orbits of an
$m$-armed spiral lie where
\begin{equation}
m(\Omega_p - \Omega) = l\kappa,
\label{eq.res}
\end{equation}
with $\Omega \equiv V_0/R$ being the angular frequency of a circular
orbit, $\kappa = \surd2V_0/R$ the epicyclic frequency in the Mestel
disc, and $l=0,\pm1$. The corotation resonance has $l=0$, $l=-1$ for
the inner, and $l=+1$ for the outer, Lindblad resonance.  The Lindblad
resonances approximately limit the radial extent of spiral modes.  For
an $m=2$ disturbance having $\Omega_p=0.141$ (this case), $R_{\rm
  CR}=7.09$, $R_{\rm ILR}=2.08$, and $R_{\rm OLR}=12.1$.

\begin{figure*}
\includegraphics[width=.8\hsize,angle=0]{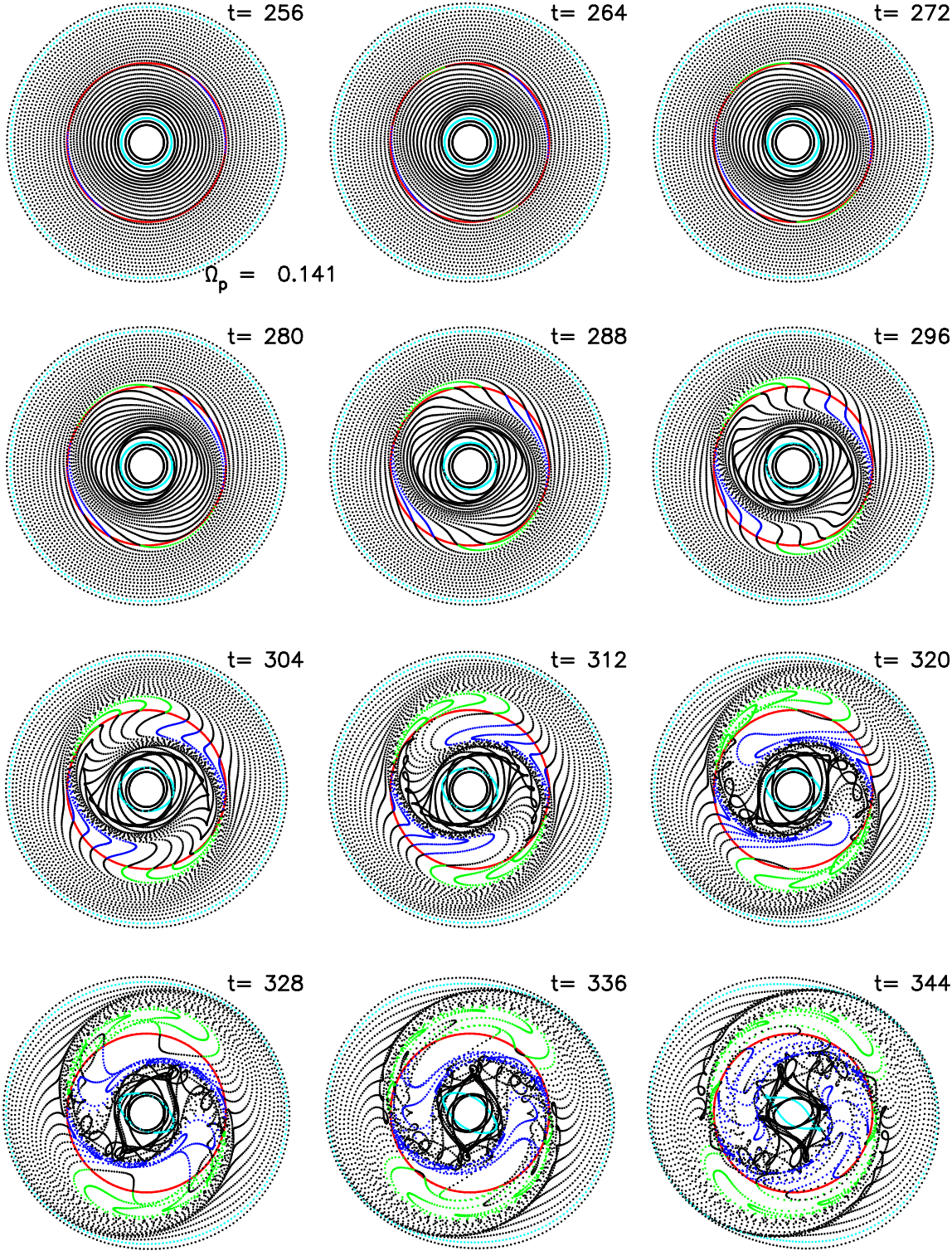}
% jasmine:/data/sellwood/5200/5205/rings2.s
\caption{Snapshots showing the positions of the test particles, that
  had initially circular orbits, over a short time interval around the
  time that the spiral instability saturates. The red circle marks the
  corotation radius of the linear instability, which has the estimated
  pattern speed $\Omega_p=0.141$, and the model rotates
  counter-clockwise.  Each panel is rotated through an angle
  $-\Omega_pt$ in order that the orientation of the disturbance
  changes little from frame to frame.  Green (outward movers) and blue
  (inward movers) particles have crossed the corotation resonance
  since the start, those that are black are the same side as at the
  start. The two rings highlghted in cyan are those closest to the
  Lindblad resonances, and remain only mildly distorted until well
  after the mode saturates.}
\label{fig.rings2}
\end{figure*}

\subsection{Orbit deflections}
\label{sec.rings2}
Figure~\ref{fig.rings2} shows the evolution of rings of test particles
that spanned the radial range of the linear instability.  The red
circle marks the radius of corotation at $R_{\rm CR} = 7.09R_0$, while
the cyan colored particles are those in the two rings each closest to
the Lindblad resonances.  Each panel is rotated through an angle
$-\Omega_pt$ in order that the orientation of the disturbance changes
little from frame to frame.  The sense of rotation of the disc is
counter-clockwise.  Particles that have crossed the corotation
resonance (CR) since the start are colored green or blue, other
particles have not crossed the CR.

Although the instability has been growing exponentially from the
start, its amplitude at $t=256$ is still low and has created barely
noticeable distortions of the rings.  Inside the CR, the orientation
of the mild oval distortions shifts with radius, creating the spiral
density concentrations.  Notice that the rings near the Lindblad
resonances, coloured cyan, remain only mildly oval until after
$t=300$.  Larger distortions near corotation become apparent after
$t=280$ that differ for particles at differing phases relative to the
growing and rotating disturbance.  Some particles have moved inwards,
and have begun to run ahead of their unperturbed motion.  At the same
time, we also see that other particles have risen outwards and begun
to lag behind.  This behaviour is accounted for as follows.

Any particle, whether massive or not, near to corotation moves slowly
with respect to the spiral perturbation and therefore experiences
almost steady forcing from the wave.  Particles orbiting just behind a
density excess are attracted forward by it and therefore gain angular
momentum.  However, the consequence of gaining significant angular
momentum from a large-amplitude disturbance is that the particle moves
onto an orbit having a larger guiding centre radius, and its angular
frequency about the galaxy centre therefore decreases.  Conversely,
particles just ahead of the perturbation are pulled back, lose angular
momentum and sink inwards, where they orbit at higher frequency.
These changes to the guiding centre radius may cause the particle to
cross corotation, and therefore to reverse its sense of motion
relative to the disturbance, as shown by the coloured particles in
Figure~\ref{fig.rings2}.  This is, of course, part of the explanation
for the phenomenon of radial migration \citep{SB02}.  The distance
from corotation within which the anomalous orbit reversals in the
corotating frame occur, which for a steady perturbation is known as
the Hill radius \citep{BT08}, grows with the amplitude of the spiral
potential and is negligible at small amplitude.

Note that the behaviour in Figure~\ref{fig.rings2} is simplified
because the unperturbed orbits were circular and disturbance forces
arose from only a single sectoral harmonic.  The net response of the
massive particles that generally have non-circular orbits would be
blurred by their random motion, but the guiding centres of those
orbits would follow the pattern illustrated, at least for those having
small initial epicycles.

\subsection{Saturation mechanism}
\label{sec.mechanism}
For a linearly unstable mode, the first-order orbit deflections due to
the growing disturbance potential create a density response that gives
rise to the perturbing potential, which is the necessary
self-consistency condition for a mode.  First order changes to orbits
in the disc may be computed as mild departures from their unperturbed
orbits \citep[\eg][]{Kaln71, BT08} for as long as the disturbance
amplitude remains small.  As the amplitude rises, however, the
approximation that the perturbed orbits can be computed as small
departures from their unperturbed orbits breaks down, because either
the perturbations become too large for first order theory to be
adequate, or changes to the unperturbed orbits can no longer be
ignored, or both.

Two examples of changes to the unperturbed orbits are: (a) a bar
forming mode saturates when the orbits become trapped by the bar
potential as its amplitude rises, and (b) the amplitude reaches that
at which orbit distortions become large.  In either case, the
self-consistency requirement of an exponentially growing mode breaks
down and the instability saturates.  This is borne out for the present
case by the evidence in Figure~\ref{fig.rings2}; linear growth of the
mode ends around time 300 (Figure~\ref{fig.amplot}), which is when the
shapes of the orbits near the CR begin to depart from the simple
elliptical distortions that would be predicted by linear perturbation
theory.  Orbit reversals in the rotating frame, which can be thought
of as a slightly different form of trapping \citep{SB02}, cause the
density response to the growing potential to disperse, whereas prior
to their onset the mild deflections reinforced the growing mode.

\begin{figure}
\includegraphics[width=\hsize,angle=0]{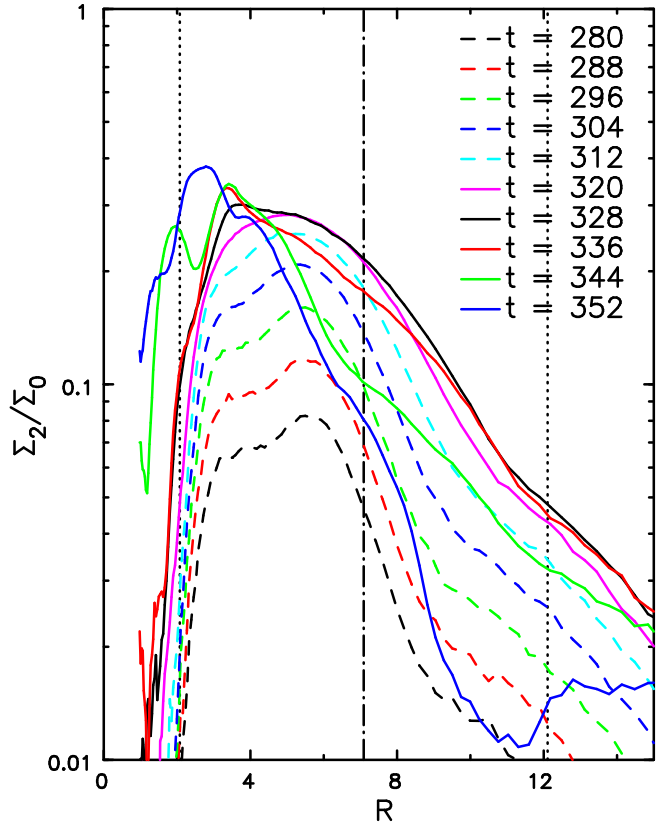}
% jasmine:/data/sellwood/5200/5205/danl.s
\caption{The time evolution of the relative overdensity of bisymmetric
  distortions of the surface density in the later stages of the
  simulation after the first mode has saturated.  The dashed curves
  show the mode amplitude still growing, although the growth rate is
  beginning to slow by $t=304$.  The vertical dash-dot line marks the
  corotation radius, and the dotted lines the radii of the Lindblad
  resonances.  The other curves are described in the text.}
\label{fig.danl}
\end{figure}

Figure~\ref{fig.danl} displays the relative overdensity of bisymmetric
distortions of the surface density at times during the later stages of
the simulation, and the radii of the principal resonances of the first
mode are marked.  The dashed curves, which peak somewhat inside
corotation, show the amplitude still growing, although the growth rate
is beginning to slow by $t=296$.  Note also that the linear mode
amplitude drops precipitously towards the ILR.  Thus for the groove
mode, the important process that causes the instability to saturate is
the change in the shapes of the orbits near to corotation, which
happens over a short period (Figure~\ref{fig.rings2}).

\citet[][their \S8.3.1]{BT08} find that the Hill radius for a steady
perturbing potential varies as the cube root of the perturber mass, or
amplitude and is therefore tiny as the mode starts to grow.  In our
case, the disturbance is growing exponentially, but the amplitude
dependence they find may still account for the sudden change of
behaviour we illustrate for the test particles in
Figure~\ref{fig.rings2}.  The massive particles are similarly
affected, as we show in \S\ref{sec.migrat}, which causes the
instability to saturate.  This idea was outlined by \citet{SB02} and
restated by \citet{SC14}, but without supporting evidence, such as
that shown in Figures~\ref{fig.amplot}, \ref{fig.rings2}, \&
\ref{fig.danl}.

Notice the pronounced amplitude contrast between the spiral in the
inner and outer disc in our Figure~\ref{fig.danl}, which was also
evident in the ``dust-to-ashes'' figure in \citet{Toom81}.  In the
absence of an external torque on the disc, the angular momentum
content, or wave action, of the two parts of the wave must be equal,
though of opposite sign.  \citet{SM22} argued that conservation of
wave action in the shrinking geometric area as the wave propagates
inwards keeps the amplitude of the spiral high, whereas conversely the
spiral amplitude diminishes as it spreads outward beyond corotation.

\begin{figure}
\includegraphics[width=.95\hsize,angle=0]{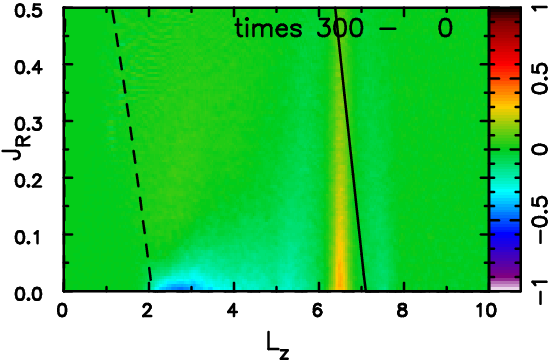}
% jasmine:/data/sellwood/5200/5205/actdiff.s 0 300
\includegraphics[width=.95\hsize,angle=0]{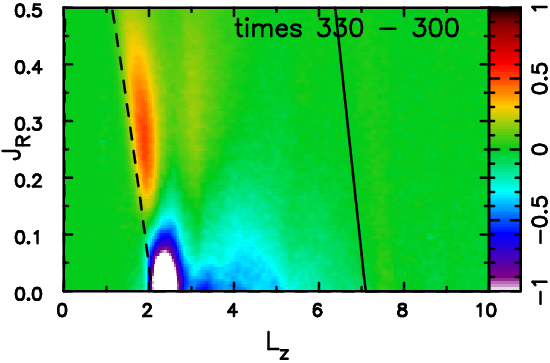}
% jasmine:/data/sellwood/5200/5205/actdiff.s 300 330
\includegraphics[width=.95\hsize,angle=0]{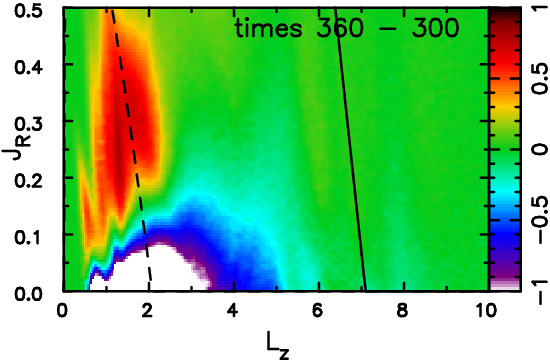}
% jasmine:/data/sellwood/5200/5205/actdiff.s 300 360
\caption{Changes to the relative density of particles in the space of angular
  momentum and radial action between times 0 and 300 (top), times 300
  and 330 (middle) and times 300 and 360 (bottom).  The solid line
  marks the locus of the corotation resonance of the first mode, the
  dashed line that of its ILR.  Notice that although the first mode
  had begun to saturate by $t=300$, strong scattering at the ILR did
  not occur until later.  The most evident change in the top panel is
  simply that the initial groove, which was inside corotation has
  filled in. \Ignore{ At the later time, additional scattering has
    occured at smaller $L_z$ because a new mode was excited, as
    reported by \citet{SB02}.}}
\label{fig.actdifs}
\end{figure}

\subsection{Subsequent evolution}
Figure~\ref{fig.amplot} shows the growth rate eases at $t\sim300$, but
both there and in Figure~\ref{fig.danl} the disturbance density over
much of the radial range continues to rise to $t\simeq320$.  Later, to
$t=352$ the wave amplitude continues to increase well interior to
corotation, while decreasing toward and beyond the CR, providing some
evidence that the disturbance is travelling away from the CR at late
times.

Linear theory \citep{Toom69} predicts that a trailing spiral wave
packet is carried away from corotation, both inward and outward, at
the group velocity.  It travels radially until it is absorbed at a
Lindblad resonance \citep{LBK}.  The absorption of the first
disturbance at its ILR is partly obscured in Figure~\ref{fig.danl} by
the new vigorous instability that develops in the inner disc, which is
probably the reason that the last two curves in this Figure show the
disturbance density rising again near the ILR.

In fact, the spiral persists for some time and has not fully decayed
even by $t\sim352$, which is more than one full turn of the
disturbance after saturation ($2\pi/\Omega_p = 43.6$ dynamical times).
A second instability, also reported by \citet{SB02}, is growing at
this time, and superposition of two or more waves causes the
transforms of the entire disc, shown in Figure~\ref{fig.amplot}, to
appear incoherent from then to the end of the simulation.

\begin{figure}
\includegraphics[width=.95\hsize,angle=0]{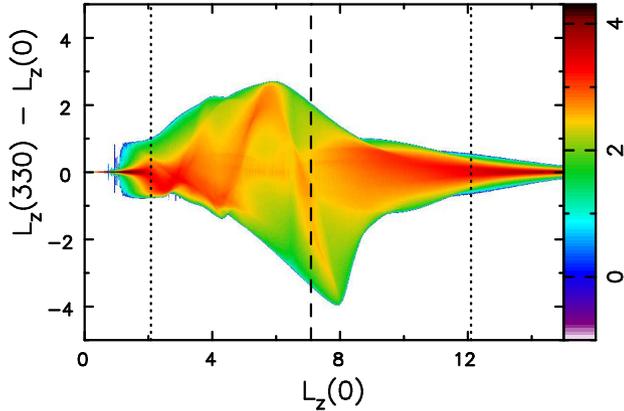}
% jasmine:/data/sellwood/5200/5205/migrat.s 330
\caption{Changes to the angular momenta of the massive particles
  between times 0 and 330.  Radial migration is again clear in this
  rerun of the case reported by \citet{SB02}, but here with 100 times
  more particles.  The dashed line marks the angular momentum of
  circular orbits at corotation, the dotted lines the Lindblad
  resonances.  The colour scale shows the logarithm of the number of
  particles in each pixel.}
\label{fig.migrat}
\end{figure}

\subsection{Resonant scattering}
Scattering of particles at the Lindblad resonances is second order in
the perturbation amplitude \citep{LBK}, and therefore starts as the
instability begins to grow from very low amplitude.  But here we show
that most scattering occurs after the mode has saturated.  This is
because the instability has acquired a store of angular momentum
\citep{LBK} that is carried away from corotation at the group velocity
\citep{Toom69}.

The radii of the principal resonances for circular orbits were defined
in eq.~(\ref{eq.res}) and evaluated there for this mode.  The angular
momenta of more general non-circular orbits that are also in resonance
decrease as the orbits become more eccentric.

Figure~\ref{fig.actdifs} shows changes to the density of particles in
action space over the time interval from the start to when the mode
saturates (top panel) and two short periods thereafter (middle and
bottom panels).  The contrast between the top and the other two panels
makes it clear that most scattering at the ILR occurs while the mode
decays, as the angular momentum stored in the disturbance drains onto
the resonance.  Note that the changes at the ILR in the middle panel
are not perfectly aligned with the resonance line and there is an
apparent weaker scattering feature at $L_z=3$ and large $J_R$.  Since
we computed actions assuming an axisymmetric potential, both these
discrepancies from expectations may be due to the continued presence
of the spiral.  In particular, they disappear by $t=360$ when the
spiral is weaker (bottom panel) although additional features at small
$L_z$ associated with the second mode have appeared by then.

While \citet{SC19} convincingly demonstrated that scattering at a
Lindblad resonance seeded a new instability, it is clear from this
discussion that the new instability may already have begun to grow
before the process of resonance scattering has completed.

\begin{figure}
\includegraphics[width=.95\hsize,angle=0]{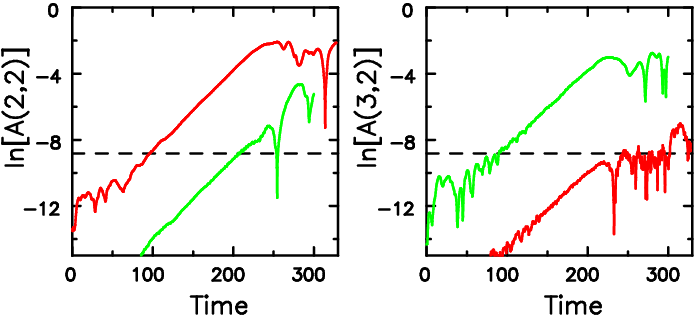}
% jasmine:/data/sellwood/5200/5221/analys.s
\caption{The amplitude evolution of a trailing $m=2$ coefficient on
  the left and a trailing $m=3$ coefficient on the right in two
  simulations.  The red lines are from a case in which the $m=2$ seed
  amplitude was enhanced while $m=3$ seed amplitude was enhanced for
  the results reported by the green lines.  In both cases, the
  disturbance forces included multiple sectoral harmonics, and the
  linear growth of modes having comparable growth rates can be seen in
  both panels.}
\label{fig.twomodes}
\end{figure}

\subsection{Radial migration}
\label{sec.migrat}
It is remarkable that there is no strong feature in any of the panels
of Figure~\ref{fig.actdifs} near corotation at $L_z=7.09$, which
should be where radial migration is strongest \citep{SB02}.  The top
panel shows only that the initial groove, which is inside corotation
for the mode, has filled, leaving faint deficiencies on either side.
Notice that corotation for the mode is at a larger radius than the
groove, even though \citet{SK91} predicted it should lie at the groove
centre; this difference is due to curvature in the global model, which
was neglected in their local analysis.

\begin{figure*}
\includegraphics[width=.65\hsize,angle=0]{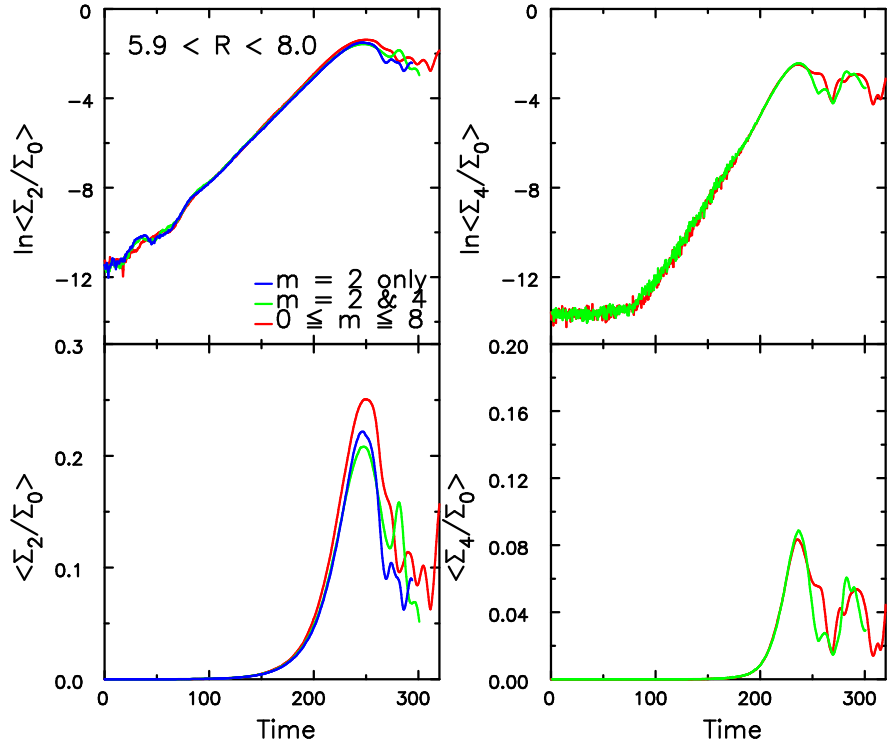}
% jasmine:/data/sellwood/5200/5223/analysd.s
\caption{The amplitude evolution of the relative disturbance density,
  averaged over the given radial range that includes corotation of the
  $m=2$ mode.  The upper panels use a logarithmic scale for the
  amplitude, while the lower panels show the same data on a linear
  scale; the left panels report the $m=2$ amplitude and the right
  panels the $m=4$ amplitude from the same simulations.  The time axis
  applies to the simulation shown by the red curve, while the other
  cases were shifted in time in order that they pass through the same
  point in the middle of the linear growth.  Disturbance forces in the
  three simulations, in which the $m=2$ seed amplitude was enhanced,
  were computed using the indicated numbers of sectoral harmonics in
  the left panels, and the same colours are used in the right hand
  panels, although there can obviously be no data in those panels from
  the $m=2$ only case.}
\label{fig.mtwo}
\end{figure*}

\begin{figure*}
\includegraphics[width=.65\hsize,angle=0]{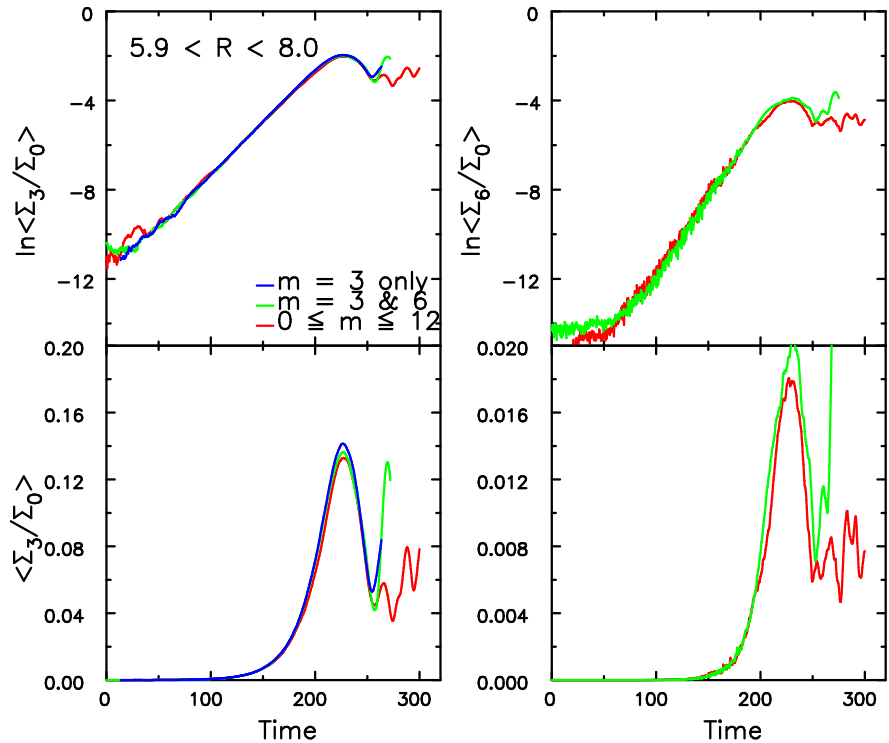}
% jasmine:/data/sellwood/5200/5225/analysd.s
\caption{As for Figure~\ref{fig.mtwo} but from simulations that
  started with the $m=3$ amplitude enhanced.}
\label{fig.mthree}
\end{figure*}

\begin{figure*}
\includegraphics[width=.8\hsize,angle=0]{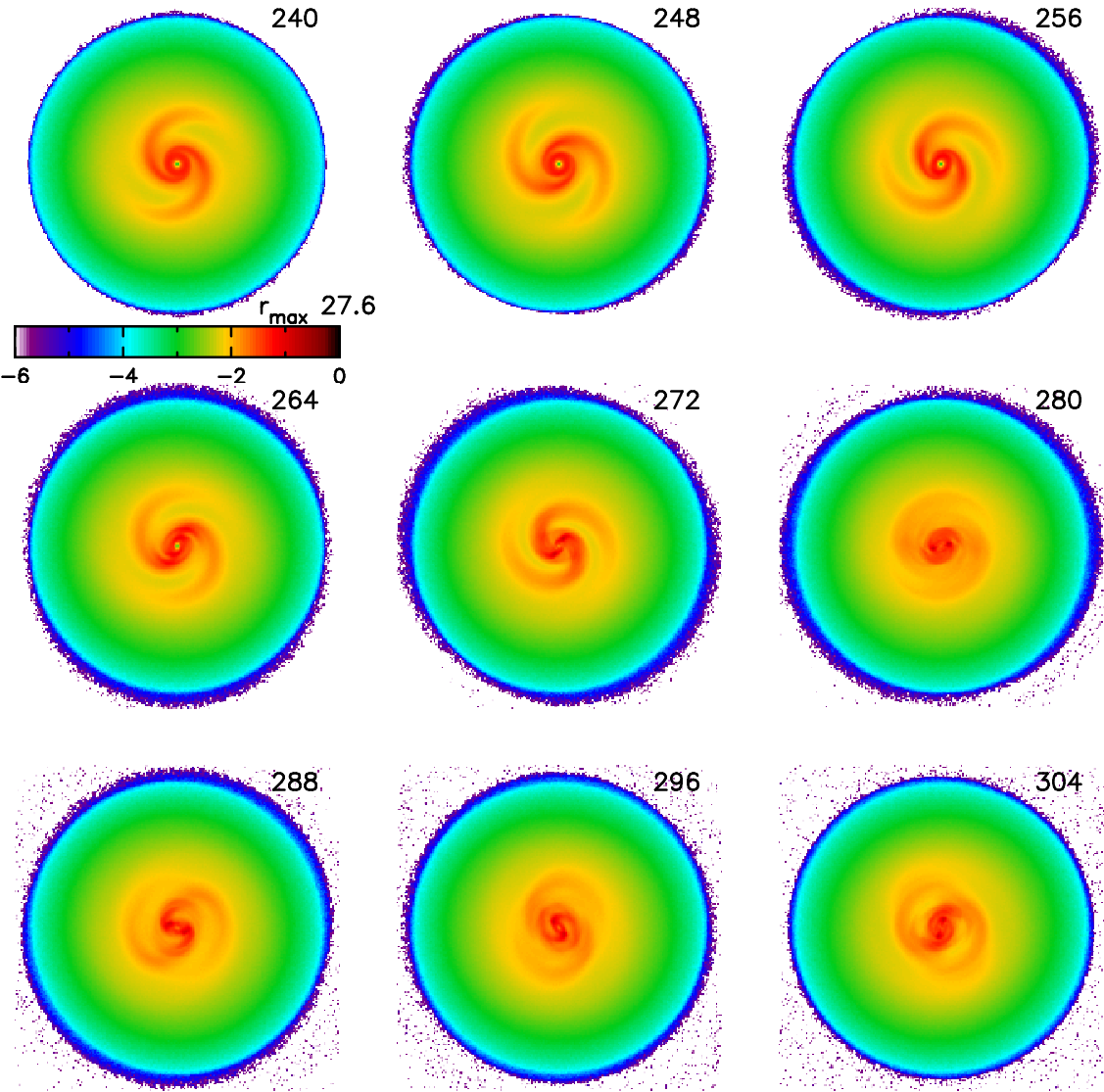}
% jasmine:/data/sellwood/5200/5221/proj.s
\caption{Snapshots showing the logarithm of the disc surface density
  over the time interval as the mode saturates ($t\sim250$) and
  continuing to much later.  Notice that the spiral in the disc
  persists well after the mode has saturated, and has not fully
  decayed even by $t=304$.  This simulation employed sectoral
  harmonics $0 \leq m \leq 8$, and the $m=2$ seed amplitude was
  boosted.}
\label{fig.proj}
\end{figure*}

Figure~\ref{fig.migrat} confirms that most massive particles close to
corotation have in fact suffered large changes to $L_z$ by $t=330$, a
time when the spiral is fading, but still visible.  These changes were
also illustrated by the behaviour of the test particles in
Figure~\ref{fig.rings2}.  The absence of a corresponding feature in
Figure~\ref{fig.actdifs} reflects the fact that radial migration
really does not heat the disc and the guiding centres of orbits are
simply interchanged.  Note that the changes at the Lindblad resonances
in Figure~\ref{fig.migrat} are much smaller than those at corotation.

\subsection{Including other sectoral harmonics}
\label{sec.general}
While instabilities having different rotational symmetries are
decoupled at small amplitude, that ceases to be true at large
amplitude.  As exemplified in Fig.~\ref{fig.rings2}, orbit deflections
ceased to be sinusoidal even as the forcing terms, in this case,
remained so.  Therefore, disturbances at different $m$ become coupled
at large amplitude, and a full description of the non-linear behaviour
requires including multiple sectoral harmonics, as we now describe.

A slight complication is that our adopted model possesses
instabilities having similar growth rates at both $m=2$ and $m=3$, as
shown in Figure~\ref{fig.twomodes}.  The growth rate of groove modes
is enhanced by the disc supporting response, which is directly related
to swing-amplification (see \S\ref{sec.mechnsm}).  In our half-mass
Mestel disc model, the swing-amplification parameter $X=2$ for $m=2$
and $X=4/3$ for $m=3$, and these values straddle the amplification
peak \citep{Toom81, BT08} causing the two modes to grow at very
similar rates.  The red lines in this figure are from a simulation in
which we perturbed the quiet start to enhance the seed amplitude at
$m=2$, while the green lines show the consequence of enhancing the
$m=3$ seed amplitude.  This trick enables us to study the saturation
of each mode separately, while the growing amplitude of the second
mode remains sufficiently low that it has a negligble effect on the
saturation of the dominant mode.  Disturbance forces in both these
simulations included multiple sectoral harmonics: $0 \leq m \leq 8$
for the case with $m=2$ enhanced and $0 \leq m \leq 12$ for the other
case.  We have confirmed by fitting modes, that the {\em linear}
growth rate, as is also evident from this Figure, and the pattern
speed of each mode is unaffected by the amplitude of the other.

However, the non-linear evolution is mildly changed when multiple
sectoral harmonics contribute to the disturbance forces, as
illustrated in Figures~\ref{fig.mtwo} and \ref{fig.mthree}.  The upper
left panels of both figures reconfirm that the linear growth of a mode
is unaffected by the inclusion of other harmonics, but the saturation
amplitude is mildly affected, as is more clearly illustrated by the
linear scale in the lower panels.  The RH panels reveal that the mode
in the LH panels drives a response at twice the angular periodicity,
which does not begin at first, but kicks in after a little while.
This driven response has identically the same pattern speed as the
main mode and rises exponentially at a steeper rate to saturate at
about the same time as the driving mode.  This behaviour is almost
identical in both the green and red lines, even though the red curve
is from a simulation in which many sectoral harmonics contributed to
the disturbance forces and the green curves just two. In fact, apart
from the linear instability whose seed amplitude was depressed
(Figure~\ref{fig.twomodes}), we could find no evidence for strongly
growing signal at any $m$ other than that shown in each of these
Figures.

The lower LH panels indicate a mild variation in the saturation
amplitude of the dominant mode, which is likely caused by the driven
responses illustrated on the right.  We have not found any evidence
that the peak amplitude varies systematically with the addition of
extra forcing terms in the six models shown in these two figures.  We
therefore suspect that the differences are due to the relative phases
of the dominant mode and of the driven response.  Note that the $m=6$
driven response peaks at a significantly lower amplitude
(Fig.~\ref{fig.mthree}) than the corresponding $m=4$ response
(Fig.~\ref{fig.mtwo}), and that the saturation amplitude spread of the
dominant mode is much less.

If this hypothesis is correct, it adds a degree of randomness to the
saturation amplitude, although the spread remains small.

\begin{table}
\caption{Mode frequencies (with $N=6$ million)}
\label{tab.modes}
\begin{tabular}{@{}cccl}
$m$ & $Q$ & $w_L$ & $\omega = m\Omega_p + i\beta$ \\
\hline
2 & 1.5 & 0.2  & $0.282\pm0.000 + (0.046\pm0.001)i$ \\ %5203
2 & 1.8 & 0.2  & $0.283\pm0.000 + (0.033\pm0.001)i$ \\ %5207
2 & 1.2 & 0.2  & $0.282\pm0.001 + (0.064\pm0.001)i$ \\ %5206
2 & 1.5 & 0.3  & $0.283\pm0.001 + (0.049\pm0.001)i$ \\ %5208
2 & 1.5 & 0.1  & $0.283\pm0.000 + (0.039\pm0.000)i$ \\ %5209
2 & 1.5 & 0.05 & $0.285\pm0.000 + (0.029\pm0.000)i$ \\ %5211
3 & 1.5 & 0.2  & $0.447\pm0.001 + (0.047\pm0.001)i$ \\ %5210
\hline
\end{tabular}
\end{table}

\subsection{Complete evolution}
Figure~\ref{fig.proj}, taken from the case with multiple active
sectoral harmonics ($0 \leq m \leq 8$) and the seed amplitude of $m=2$
slightly boosted, reveals that the spiral disturbance persists for
some time after the linear mode has saturated.  The second $m=2$
instability, caused by ILR scattering of the first mode, may be seen
in the inner parts.  This behaviour is barely distinguishable from
that in an $m=2$ only simulation, and it seems that inclusion of
additional harmonics has at most a very minor effect on the overall
evolution.

\section{Other models}
\label{sec.variants}
We have experimented with changing the properties of the disc, of the
groove, and changing the active sectoral harmonic.  Note that all the
experiments reported in this section employ disturbance forces that
are restricted to a single sectoral harmonic: mostly $m=2$, but $m=3$
in one case.  The measured saturation amplitude therefore avoids the
random element caused by the influence of the driven response
(\S\ref{sec.general}).  Also there is no need to enhance the seed
amplitude, and we employ the standard quiet start \citep{Sell83}.

The fitted eigenfrequencies of the linear mode in each case are given
in Table~\ref{tab.modes}.  Notice that the fitted frequency in the first
case listed in the table, for which $N=6$ million particles, is in
near perfect agreement with that when $N=60$ million
(\S\ref{sec.res1}) and when $N=0.6$ milion \citep{SB02}.

\begin{figure}
\includegraphics[width=\hsize,angle=0]{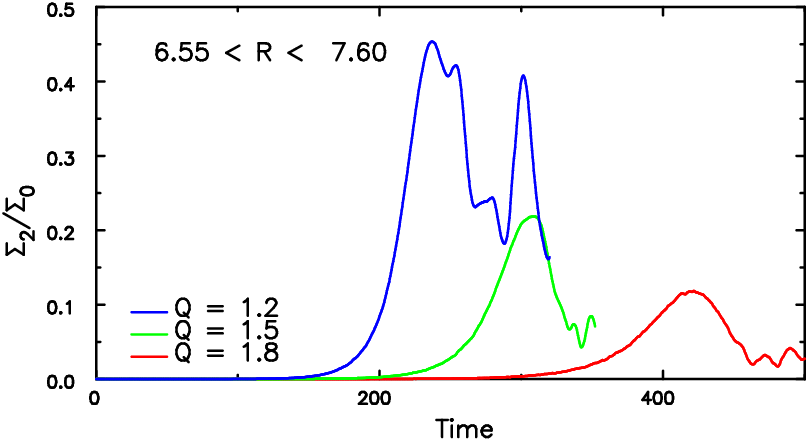}
% jasmine:/data/sellwood/5200/5207/analys-Q.s
\includegraphics[width=\hsize,angle=0]{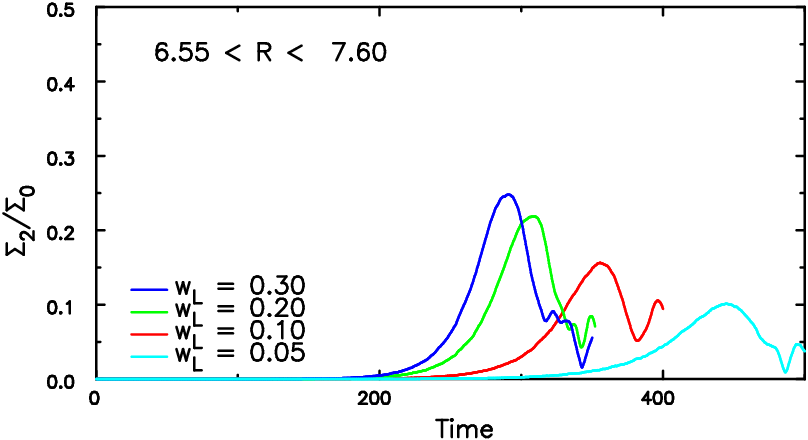}
% jasmine:/data/sellwood/5200/5207/analys-w.s
\includegraphics[width=\hsize,angle=0]{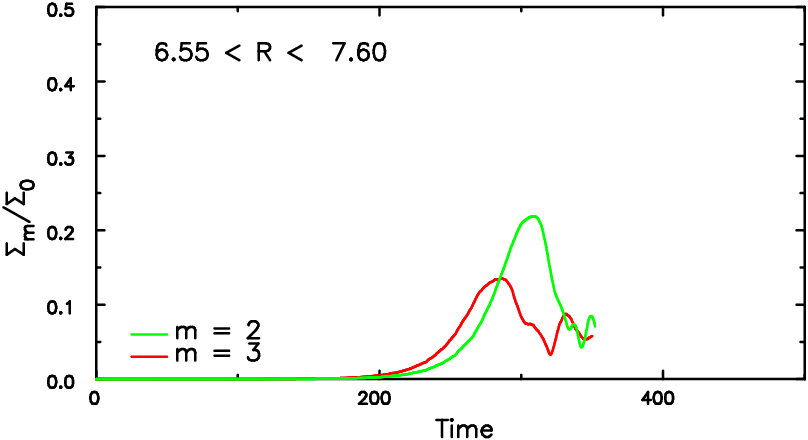}
% jasmine:/data/sellwood/5200/5207/analys-m.s
\caption{The time evolution of the relative perturbed surface density,
  averaged over the given radial range, in a series of models.  In the
  top panel, the $Q$ value for the Mestel disc is varied while the
  groove parameters are held constant, in the middle panel, the width
  of the groove is changed while $Q$ is held constant, and in the
  bottom panel, the sectoral harmonic is varied, while the groove
  parameters and $Q$ are held constant.  The green curve in all three
  panels is from the same simulation.}
\label{fig.amp-vars}
\end{figure}

\subsection{New findings}
Figure~\ref{fig.amp-vars} reports the time evolution of the mean
relative amplitude of the disturbance density around the radius of
corotation.  The top panel shows the effect of changing the nominal
value of $Q$ of the Mestel disc, while keeping the groove parameters
unchanged.  It is clear that both the growth-rate and saturation
amplitude of the groove mode decrease with increasing $Q$; the pattern
speed, which is strongly tied to the circular angular frequency at the
radius of the groove, remains nearly the same, however
(Table~\ref{tab.modes}).  The growth rate is enhanced by the
supporting response from the disc surrounding the groove, as described
in \S\ref{sec.mechnsm}, which becomes more vigorous with decreasing
$Q$.

The middle panel of Figure~\ref{fig.amp-vars} shows the effect of
varying the groove width, while holding $Q=1.5$. As expected from
theory for narrow grooves \citep{SK91}, the growth rate of the mode
increases with the width of the groove.  Once again we see that the
saturation amplitude varies with the growth rate of the mode.

The bottom panel of Figure~\ref{fig.amp-vars} shows the effect of
changing the active sectoral harmonic from $m=2$, which is the default
for all other simulations, to $m=3$.

\begin{figure}
\includegraphics[width=\hsize,angle=0]{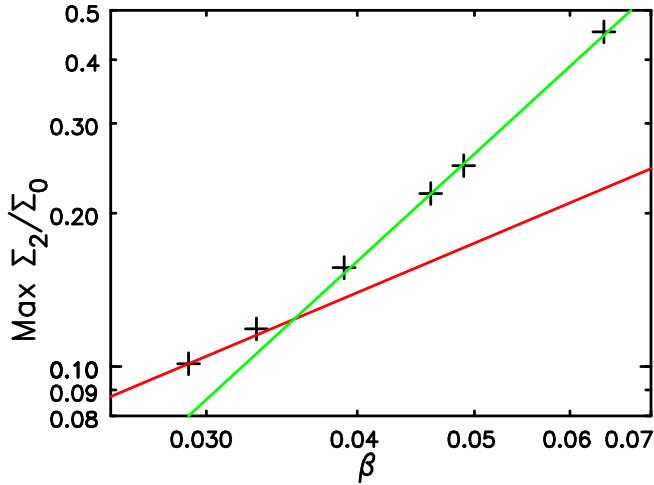}
% herschel1:/home/sellwood/papers/SC22/ampgrw.s
\caption{The growth rate dependence of the saturation amplitude of the
  $m=2$ modes shown in the top two panels of
  Figure~\ref{fig.amp-vars}. The points do not fit a straight line
  even with logarithmic scaling.  The red and green lines, which have
  slopes 1 and 2, respectively are not fits to the points and are
  drawn only to guide the eye.}
\label{fig.ampgrw}
\end{figure}

\subsection{Dependence on growth rate}
The top two panels of Figure~\ref{fig.amp-vars} indicate that the
saturation amplitude of the mode is lower for more slowly growing
modes.  In fact, Figure~\ref{fig.ampgrw} reveals quite a tight
correlation for these six cases.  Note that our measured growth rates,
$\beta$, have small uncertainties, but we have no information about
possible uncertainties in the limiting amplitude, which would have
required us, for example, to have run ensembles of simulations for
each case having different random seeds.  Note also that had we
included other sectoral harmonics, the amplitude peaks may have been
slightly affected in an unpredictable manner, as described in
\S\ref{sec.general}, making the correlation less tight.  However, as
we have often found, the simpler physics in these restricted cases
enables us to learn more.

The logarithmic scaling in Figure~\ref{fig.ampgrw} indicates that the
functional relation between growth rate and saturation amplitude
dependence is not a simple power law, which we have been unable to
explain.  However, the referee suggested that the saturation amplitude
should vary as $\beta^2$, which appears to fit the more vigorous
instabilities but is inconsistent with the two most slowly growing
modes.  The argument was based on the expected libration times of
orbits trapped at the CR, which unfortunately would be best defined
for the most slowly growing modes.

\begin{figure*}
\includegraphics[width=.8\hsize,angle=0]{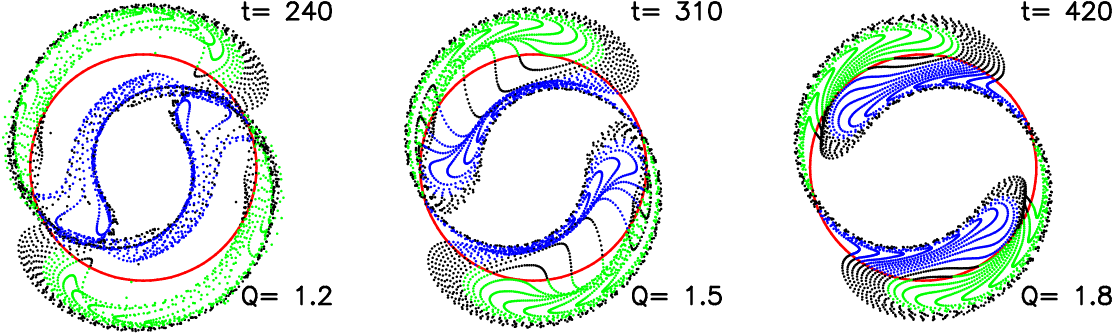}
% jasmine:/data/sellwood/5200/5206/rings.s
\caption{Test particle rings that began in the vicinity of CR only in
  the three simulations having different $Q$.  The time displayed is
  when the instability first saturates in each case.  Notice that the
  radial displacements are quite similar in all three cases, even
  though saturation occurs at very different amplitudes and times.}
\label{fig.ringsQ}
\end{figure*}

\begin{figure}
\includegraphics[width=\hsize,angle=0]{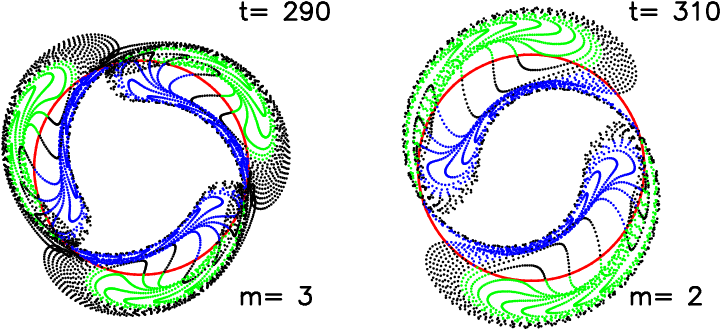}
% jasmine:/data/sellwood/5200/5210/rings.s
\caption{As in Figure~\ref{fig.ringsQ}, but for the two simulations
  having different $m$.  The time displayed is when the instability
  first saturates in each case.  Notice in the left-hand panel that
  particle displacements from different spiral arms are beginning to
  overlap.}
\label{fig.rings-m}
\end{figure}

We offer the following qualitative explanation for the interdependence
of growth rate and saturation amplitude in terms of the saturation
mechanism proposed above (\S\ref{sec.mechanism}).  There we argued
that growth was halted by orbit deflections that are particularly
large near the corotation radius of the mode because particles drift
slowly relative to the perturbation.  In fact, it is reasonable that
the saturation amplitude should increase as the growth rate increases
because the perturbation acts on the orbits for a shorter period.
Since changes to the unperturbed orbits depend on {\em both} the
amplitude of the pertubation and the length of time that it acts,
disturbances reach higher amplitudes for vigorous instabilities than
for those from more slowing growing modes, consistent with our
results.

The distortions to the rings of test particles at the times the mode
saturates are quite similar in the three cases having different $Q$
(Figure~\ref{fig.ringsQ}), consistent with this being a threshold
distortion that is reached at lower amplitude because the perturbing
forces act for longer when the mode grows slowly.  A more careful
examination of the Figure shows that radial displacements are slightly
smaller for the more slowly growing mode (RH panel) and greater for
the rapidly evolving case (LH panel) and these differences become more
pronounced at later times.

The quasi-steady density-waves proposed by \citet{BLLT}, if they occur
at all, have a different mode mechanism, and therefore may not
saturate in the same manner.  However, the tight relation revealed in
Figure~\ref{fig.ampgrw} suggests that the very slowly growing modes
they favour may saturate at a tiny amplitude.  This would be because
large orbit deflections in the vicinity of corotation will inevitably
build up over the long growth time, leading to the breakdown of the
linear approximation and mode saturation at small amplitude.

The different saturation amplitudes for changes to the sectoral
harmonic reported in the bottom panel of Figure~\ref{fig.amp-vars}
cannot, however be attributed to growth rate differences.  In fact,
the instabilities at $m=2$ and $m=3$ grow almost equally fast
(Table~\ref{tab.modes}) in the linear regime, for the reason given in
\S\ref{sec.general}.  It seems likely that saturation amplitude
differs because the spatial scales of the spiral instabilities differ,
and growth is limited by the scale of the orbit perturbations relative
to the spatial scale of the mode, as illustrated in
Figure~\ref{fig.rings-m}.  This argument suggests that, other things
being equal, the saturation amplitude of spiral modes should also
decrease as the order of rotational symmetry increases.

\section{Conclusions}
We have presented an in-depth study of how a single unstable spiral
mode in a simplified model ceases to grow and the manner in which it
decays.  The end of exponential growth is quite abrupt
(Figure~\ref{fig.amplot}), and is associated with large changes to the
orbit guiding centres in the vicinity of corotation
(Figure~\ref{fig.rings2}) that destroy the linear density response.
These orbit deflections, which do not increase random motion
(Figure~\ref{fig.migrat}), are also responsible for radial
migration \citep{SB02}.

The spiral fades gradually after it saturates (Figure~\ref{fig.proj}),
persisting for over a full turn while the radial dependence of its
amplitude changes (Figure~\ref{fig.danl}) as it travels inward inside
corotation, as expected \citep{Toom69}.  The wave possesses a store of
angular momentum that was built up as it grew, but since the net
angular momentum of an isolated disc cannot change, the mode has
reduced the angular momentum of the inner disc and raised that of the
outer \citep{LBK}.  The negative angular momentum stored in the wave
drains onto particles at the inner Lindblad resonance, which gain
energy and radial action while losing angular momentum.  It is
noteworthy that we found almost all scattering occurs after the wave
has started to decay (Figure~\ref{fig.actdifs}).

We reported results for other groove modes in \S\ref{sec.variants},
finding (Figure~\ref{fig.ampgrw}) a surprsisingly tight correlation
between the saturation amplitude and the linear growth rate of the
mode, in the sense that the maximum amplitude is lower for
instabilities that grow more slowly.  This is because the more gradual
growth of slowly growing modes creates at lower amplitude the large
orbit deflections near corotation that destroy the self-consistency
requirement of a linear mode.  We suggest that this amplitude
dependence on growth rate could add a third issue, to the two
previously noted in the introduction, why the mechanism for the
quasi-steady spiral modes proposed by \citet{BLLT} seems unlikely to
be viable.

The non-linear evolution is only slightly affected by including other
sectoral harmonics (Figures~\ref{fig.mtwo} and \ref{fig.mthree}).  The
mild changes to the saturation amplitude would probably blur the
correlation in Figure~\ref{fig.ampgrw} a little, but not change the
underlying physical explanation for it.

This series of papers has presented an explanation for the origin and
properties of the spirals in simulations of unbarred, isolated, discs
that we hope readers will find convincing.  Along the way, we have
shown that other theories for self-excited spiral patterns in galaxies
have multiple issues (\S\ref{sec.intro}) and that no other theory
accounts for the preference for open, somewhat ragged, spiral patterns
having low-order rotational symmetry that is manifested by real
galaxies \citep[\eg][]{Davis12, Hart2016, Yu18}.

Whether spirals in real galaxies are self-excited by our suggested
mechanism is yet to be demonstrated, however, and a meaningful
confrontation with observational data is very hard to devise.
Observations have established that spiral patterns in real galaxies
are azimuthal variations in the surface mass density of the old disc
stars (\eg\ \citealt{Zibetti09} and the review by \citealt{SM22}), the
gravitational forces from which deflect the gas flow in a systematic
manner as it streams through the spiral \citep[\eg][]{Kranz03,
  Shetty07, Erroz15}, conclusively establishing that spirals are
density waves in the disc.

As outlined in \S\ref{sec.intro}, we argue that spirals may be the
superposition of two or more spiral modes that are excited by separate
groove instabilities, and it is this statement that is hard to test.
Note that evidence will have to be indirect, because we have a single
snapshot of each galaxy that is not expected to change significantly
for millions of years.

Assuming spiral arms trigger star formation and that stars and gas
stream through the pattern, \citet{Dixon71} and \citet{DP10} proposed
that age gradients among the stars downstream from a constant pattern
speed arm should become shallower farther from corotation.
\citet{Foyl11} rule out age gradients downstream from the spiral in
their sample, but others \citep{Chan17, YH18, Mill19, Pete19} claim to
have detected them.  However, none of these careful studies was able
to establish a fixed pattern speed over the entire radial extent of
the spiral.  \citet{PD19} also suggested that $\cot \alpha$, with
$\alpha$ being the pitch angle of the spiral, should have a uniform
distribution across some range of $\alpha$ and over spiral arms in
many galaxies if spirals wind up over time.  \citet{Li21} applied this
test to a sample of 200 galaxies, finding $\cot \alpha$ values that
were consistent with a uniform distribution over the range $15^\circ
\la \alpha \la 50^\circ$.  However, winding arms are expected
\citep{SC21} when the spiral results from the superposition of
multiple modes having distinct pattern speeds, amplitudes, and perhaps
also rotational symmetries.

While the detailed response of gas to spiral arms is an area of
current research \citep[\eg][]{KKO20}, we expect its flow to adjust
quickly to the slow changes in the spiral potential, and therefore to
reflect the instantaneous forcing pattern.  But distentangling the
projected perturbed velocity flow of this clumpy medium into its
separate contributions from the possible multiple modes, which each
have a set of resonances at different radii, presents a considerable
challenge, but one which may be possible with observational data of
sufficiently high spatial and velocity resolution.

The possible identification of multiple modes would be encouraging,
but what really needs to be tested is the mechanism that excites them,
which is to show that the distribution of stars in the disc is not a
smooth function of their angular momenta.  There is little prospect in
the foreseeable future to establish the detailed positions and
velocity components of a sufficient number stars in any external
galaxy that could reveal minor gaps in their phase space
distribution. But the \Gaia\ satellite is gathering these data for the
Milky Way.  \citet{STCCR} selected from the DR2 \citep{Gaia18} $\ga
300$K stars near the Sun having accurate data and small vertical
motions, and converted their 4D phase space information into action
angle variables, finding a non-smooth distribution in all components
but, most tellingly, the kind of scattering features expected from
resonant scattering in the distribution of angular momentum,
resembling that shown in Figure~\ref{fig.actdifs}.  New data releases
will provide more accurate phase space data for more stars, which will
be invaluable.  But the Milky Way is a strongly barred, mildly warped
galaxy having infalling satellites, so that a conclusive finding that
features in phase space are due to spiral scattering, and that have
excited spiral instabilities, will require very careful modelling.  
The Milky Way is just one galaxy, and the search for similar
confirming evidence from other galaxies is an on-going challenge.

\section*{Acknowledgements}
We thank the referee for insightful comments that have helped us to
strengthen our conclusions.  JAS acknowledges the continuing
hospitality of Steward Observatory.

\section*{Data availability}
The data from the simulations reported here can be made available
on request.  The simulation code can be downloaded from
{\tt http://www.physics.rutgers.edu/galaxy}


% Don't change these lines
\bsp	% typesetting comment
\label{lastpage}

\begin{thebibliography}{99}
\def\skip#1{ \etal\ }
\def\jcop{J. Comp. Phys.}
\def\PhD{PhD thesis.}
\def\rmp{Rev. Mod. Phys.}
\def\rpp{Rep. Prog. Phys.}

\bibitem[Aumer \etal(2016)]{ABS}
Aumer, M., Binney, J. \& Sch\"onrich, R. 2016, \mnras, {\bf 459}, 3326%-48

\bibitem[Bertin(2014)]{Bert14}
Bertin, G. 2014, \textit{Dynamics of Galaxies} 2nd ed. (Cambridge University Press, Cambridge)

\bibitem[Bertin \etal(1989)]{BLLT}
Bertin, G., Lin, C. C., Lowe, S. A. \& Thurstans, R. P. 1989, \apj, {\bf 338}, 78-103

\bibitem[Binney \& Tremaine(2008)]{BT08}
Binney J. \& Tremaine S. 2008, \textit{Galactic Dynamics} 2nd ed. (Princeton University Press, Princeton NJ)

\bibitem[Carlberg \& Freedman(1985)]{CF85}
Carlberg, R. G. \& Freedman, W. L. 1985, \apj, {\bf 298}, 486%-92

\bibitem[Chandar \etal(2017)]{Chan17}
Chandar, R., Chien, L.-H., Meidt, S.,\skip{ Querejeta, M., Dobbs, C., Schinnerer, E., Whitmore, B. C., Calzetti, D., Dinino, D., Kennicutt, R. C., Regan, M.} 2017, \apj, {\bf 845}, 78% (12pp)

\bibitem[Davis \etal(2012)]{Davis12}
Davis, B. L., Berrier, J. C., Shields, D. W.,\skip{ Kennefick, J., Kennefick, D., Seigar, M. S., Lacy. C. H. S. \& Puerari, I.} 2012, \apjs, {\bf 199}, 33

\bibitem[Debattista \& Sellwood(2000)]{DS00}
Debattista, V. P. \& Sellwood, J. A. 2000, \apj, {\bf 543}, 704

\bibitem[Dixon(1971)]{Dixon71}
Dixon, M. E. 1971, \apj, {\bf 164}, 411%-23

\bibitem[Dobbs \& Baba(2014)]{DB14}
Dobbs, C. \& Baba, J. 2014, \pasa, {\bf 31}, 35 % (40pp)

\bibitem[Dobbs \& Pringle(2010)]{DP10}
Dobbs, C. L. \& Pringle, J. E. 2010, \mnras, {\bf 409}, 396%-404

\bibitem[D'Onghia \etal(2013)]{DVH}
D'Onghia, E., Vogelsberger, M. \& Hernquist, L. 2013, \apj, {\bf 766}, 34 % (14pp)

\bibitem[Erroz-Ferrer \etal(2015)]{Erroz15}
Erroz-Ferrer, S., Knapen, J.~H., Leaman, R.,\skip{ Cisternas, M., Font, J., Beckman, J.~E., Sheth, K., Mu\~n oz-Mateos, J.~C., D\'\i az-Garc\'\i a, S., Bosma, A., Athanassoula, E., Elmegreen, B.~G., Ho, L.~C., Kim, T., Laurikainen, E., Martinez-Valpuesta, I., Meidt, S.~E. \& Salo, H.} 2015, \mnras, {\bf 451}, 1004-24

\bibitem[Foyle \etal(2011)]{Foyl11}
Foyle, K., Rix, H. -W., Dobbs, C. L., Leroy, A. K. \& Walter, F. 2011, \apj, {\bf 735}, 101% (11 pp)

\bibitem[Gaia collaboration(2018)]{Gaia18}
Gaia collaboration: Katz, D., Antoja, T., Romero-G\'o, M., \etal\ 2018, \aap, {\bf 616A}, 11 % (40 pp)

\bibitem[Goldreich \& Lynden-Bell(1965)]{GLB65}
Goldreich, P. \& Lynden-Bell, D. 1965, \mnras, {\bf 130}, 125%-58

\bibitem[Hart \etal(2016)]{Hart2016}
Hart, R. E., Bamford, S. P., Willett, K. W.,\skip{ Masters, K. L., Cardamone, C., Lintott, C. J., Mackay, R. J., Nichol, R. C., Rosslowe, C. K., Simmons, B. D. \& Smethurst, R. J.} 2016, \mnras, {\bf 461}, 3663%-82

\bibitem[Hockney \& Brownrigg(1974)]{HB74}
Hockney, R. W. \& Brownrigg, D. R. K. 1974, \mnras, {\bf 167}, 351%-58

\bibitem[Hohl(1971)]{Hohl}
Hohl, F. 1971, \apj, {\bf 168}, 343%-59

\bibitem[Julian \& Toomre(1966)]{JT66}
Julian, W. H. \& Toomre, A. 1966, \apj, {\bf 146}, 810%-30

\bibitem[Kalnajs(1971)]{Kaln71}
Kalnajs, A. J. 1971, \apj, {\bf 166}, 275%-93

\bibitem[Kim \etal(2020)Kim, Kim \& Ostriker]{KKO20}
Kim, W-T., Kim, C-G. \& Ostriker, E. C. 2020, \apj, {\bf 898}, 35 %(33pp)

\bibitem[Kranz \etal(2003)]{Kranz03}
Kranz, T., Slyz, A. D. \& Rix, H.-W. 2003, \apj, {\bf 586}, 143

\bibitem[Lingard \etal(2021)]{Li21}
Lingard, T., Masters, K. L., Krawczyk, C.,\skip{ Lintott, C., Kruk, S., Simmons, B., Keel, W., Nichol, R. C. \& Baeten, E.} 2021, \mnras, {\bf 504}, 3364%-74

\bibitem[Lynden-Bell \& Kalnajs(1972)]{LBK}
Lynden-Bell, D. \& Kalnajs, A. J. 1972, \mnras, {\bf 157}, 1%-30

\bibitem[Miller \etal(2019)]{Mill19}
Miller, R., Kennefick, D., Kennefick, J.,\skip{ Shameer Abdeen, M., Monson, E., Eufrasio, R. T., Shields, D. W. \& Davis, B. L.} 2019, \apj, {\bf 874}, 177% (12 pp)

\bibitem[Miller, Prendergast \& Quirk(1970)]{MPQ}
Miller, R. H., Prendergast, K. H. \& Quirk, W. J. 1970, \apj, {\bf 161}, 903%-16

\bibitem[Oort(1962)]{Oort}
Oort, J. H. 1962, in {\it Interstellar Matter in Galaxies}, ed.\ L. Woltjer (New York: Benjamin), p.~234

\bibitem[Peterken \etal(2019)]{Pete19}
Peterken, T. G., Merrifield, M. R., Arag\'on-Salamanca, A., \skip{ Drory, N., Krawczyk, C. M., Masters, K. L., Weijmans, A-M. \& Westfall, K. B.} 2019, Nature Astronomy, {\bf 3}, 178%-82

\bibitem[Pringle \& Dobbs(2019)]{PD19}
Pringle, J. E. \& Dobbs, C. L. 2019, \mnras, {\bf 490}, 1470%-73

\bibitem[Sellwood(1983)]{Sell83}
Sellwood, J. A. 1983, \jcop, {\bf 50}, 337

\bibitem[Sellwood(1989)]{Sell89}
Sellwood, J. A. 1989, in {\it Dynamics of Astrophysical Discs}, ed.\ J. A. Sellwood (Cambridge: Cambridge University Press) pp~155%-71

\bibitem[Sellwood(2011)]{Sell11}
Sellwood, J. A. 2011, \mnras, {\bf 410}, 1637%-46

\bibitem[Sellwood(2012)]{Sell12}
Sellwood, J. A. 2012, \apj, {\bf 751}, 44% (11pp)

\bibitem[Sellwood(2014)]{Sell14}
Sellwood, J. A. 2014, arXiv:1406.6606 (on-line manual: \hfil\break {\tt http://www.physics.rutgers.edu/$\sim$sellwood/manual.pdf})

\bibitem[Sellwood(2021)]{Sell21}
Sellwood, J. A. 2021, \mnras, {\bf 506}, 3018%-23

\bibitem[Sellwood \& Athanassoula(1986)]{SA86}
Sellwood, J. A. \& Athanassoula, E. 1986, \mnras, {\bf 221}, 195%-212

\bibitem[Sellwood \& Binney(2002)]{SB02}
Sellwood, J. A. \& Binney, J. J. 2002, \mnras, {\bf 336}, 785%-96

\bibitem[Sellwood \& Carlberg(1984)]{SC84}
Sellwood, J. A. \& Carlberg, R. G. 1984, \apj, {\bf 282}, 61%-74

\bibitem[Sellwood \& Carlberg(2014)]{SC14}
Sellwood, J. A. \& Carlberg, R. G. 2014, \apj, {\bf 785}, 137% (12pp)

\bibitem[Sellwood \& Carlberg(2019)]{SC19}
Sellwood, J. A. \& Carlberg, R. G. 2019, \mnras, {\bf 489}, 116%-31

\bibitem[Sellwood \& Carlberg(2021)]{SC21}
Sellwood, J. A. \& Carlberg, R. G. 2021, \mnras, {\bf 500}, 5043%-55

\bibitem[Sellwood \& Kahn(1991)]{SK91}
Sellwood, J. A. \& Kahn, F. D. 1991, \mnras, {\bf 250}, 278%-99

\bibitem[Sellwood \& Lin(1989)]{SL89}
Sellwood, J. A. \& Lin, D. N. C. 1989, \mnras, {\bf 240}, 991%-1007

\bibitem[Sellwood \& Masters(2022)]{SM22}
Sellwood, J. A. \& Masters, K. L. 2022, \araa, {\bf 60}, 73%-120

\bibitem[Sellwood \etal(2019)]{STCCR}
Sellwood, J. A., Trick, W. H., Carlberg, R. G., Coronado, J. \& Rix, H-W. 2019, \mnras, {\bf 484}, 3154%-67

\bibitem[Shetty \etal(2007)]{Shetty07}
Shetty, R., Vogel, S. N., Ostriker, E. C. \& Teuben, P. J. 2007, \apj, {\bf 665}, 1138%-58

\bibitem[Toomre(1969)]{Toom69}
Toomre, A. 1969, \apj, {\bf 158}, 899%-13

\bibitem[Toomre(1981)]{Toom81}
Toomre, A. 1981, In {\it The Structure and Evolution of Normal Galaxies}, eds. S. M. Fall \& D. Lynden-Bell (Cambridge, Cambridge Univ. Press) p.~111%-36

\bibitem[Toomre(1990)]{Toom90}
Toomre, A. 1990, in {\it Dynamics \& Interactions of Galaxies}, ed.\ R. Wielen (Berlin, Heidelberg: Springer-Verlag) p. 292%-303

\bibitem[Vogelsberger \etal(2020)]{Voge20}
Vogelsberger, M., Marinacci, F., Torrey, P. \& Puchwein, E. 2020, Nat. Rev. Phys. {\bf 2}, 42%-66

\bibitem[Yu \& Ho(2018)]{YH18}
Yu, S.-Y. \& Ho, L. C.\ 2018, \apj, {\bf 869}, 29% (13pp)

\bibitem[Yu \etal(2018)]{Yu18}
Yu, S-Y., Ho, L. C., Barth, A. J. \& Li, Z-Y. 2018, \apj, {\bf 862}, 13

\bibitem[Zibetti \etal(2009)]{Zibetti09}
Zibetti, S., Charlot, S., \& Rix, H.-W. 2009, \mnras, {\bf 400}, 1181%-98

\end{thebibliography}
\end{document}